\documentstyle[12pt]{article}
\catcode`\@=11

\@addtoreset{equation}{section}

\catcode`\@=11

\def\section{\@startsection{section}{1}{\z@}{3.5ex plus 1ex minus
   .2ex}{2.3ex plus .2ex}{\large\bf}}
\def\thesection{\arabic{section}.}

\def\appendix{\setcounter{section}{0}
        \def\thesection{APPENDIX }}

\newcommand{\beq}{\begin{equation}}
\newcommand{\eeq}{\end{equation}}
\newcommand{\bz}{{\bar z}}
\newcommand{\Imm}{{Im\,}}
\def\rref#1{(\ref{#1})}

\begin{document}

\begin{titlepage}
\newlength{\pubnumber}
\settowidth{\pubnumber}{September~1991}
\vspace{.5in}
\begin{flushright}
\parbox{\pubnumber}{
\begin{flushleft}
UCD-91-20\\
UBCTP-91-23\\
September 1991\\
\end{flushleft}
}
\end{flushright}
\vspace{.5in}
\begin{center}
{\Large\bf
Three-Dimensional Gravity\\and String Ghosts\\}
\vspace{.4in}
S.~C{\sc arlip}\\
       {\small\it Department of Physics}\\
       {\small\it University of California}\\
       {\small\it Davis, CA 95616}\\{\small\it USA}\\
\end{center}
\begin{center}
{\small and}\\
\end{center}
\begin{center}
I.~I.~K{\sc ogan}\footnote{On leave from ITEP, Moscow, USSR}\\
       {\small\it Department of Physics}\\
       {\small\it University of British Columbia}\\
       {\small\it Vancouver V6T 2A6}\\{\small\it Canada}\\
\end{center}

\addtocounter{footnote}{-1}

\vspace{.5in}
\begin{center}
\large\bf Abstract
\end{center}

It is known that much of the structure of string theory can be
derived from three-dimensional topological field theory and
gravity.  We show here that, at least for simple topologies,
the string diffeomorphism ghosts can also be explained in
terms of three-dimensional physics.

\end{titlepage}

\section{Introduction}

It has been realized for several years that much of string theory
can be explained in terms of three-dimensional topological field
theory.  The key to such a three-dimensional approach was
Witten's observation \cite{Wita} that the Hilbert space of a
Chern-Simons theory can be identified with the space of conformal
blocks of an associated two-dimensional conformal field theory.
This means that the Chern-Simons path integral on a manifold
with boundary $\Sigma$ has a natural interpretation as the
left-moving piece of a conformal field theory amplitude on
$\Sigma$, or, equivalently, as an important part of the integrand
of a string theory amplitude.

To obtain a complete amplitude for a conformal field theory, it
is necessary to combine left- and right-movers, of course.  But
this is not hard.  As a topological field theory, Chern-Simons
theory has a trivial propagator, so the transition amplitude on
a three-manifold with the topology $[0,1]\!\times\!\Sigma$ is
simply the overlap between the wave functions on the two
boundaries.  For appropriately chosen wave functions, this
overlap integral gives a modular invariant combination
of holomorphic and antiholomorphic conformal blocks, that is,
a full conformal field theory amplitude.  The physical picture is
that of a string ``thickened'' to an annulus; since Chern-Simons
theory is topological, it is only the string-like edge excitations
of the resulting membrane that contribute to the amplitudes.  It
is worth noting that additional three-dimensional degrees of
freedom can be readily introduced into this picture, since
Chern-Simons theory can be viewed as the infinite mass limit
of a dynamical topologically massive gauge theory
\cite{DJT,Kogan,CarKog}.

To understand this construction in more detail, consider a
Chern-Simons theory with semisimple gauge group $G$ and coupling
constant $k$ on a manifold $[0,1]\!\times\! T^2$.  The
Chern-Simons Hilbert space is spanned by functions
$\Psi_\lambda[A_{\bz}]$ proportional to the level $k$
Weyl-Ka{\v c} characters \cite{EMSS,MS}, which transform under
a unitary representation of the modular group.  Physical wave
functions, on the other hand, should presumably be exactly
diffeomorphism-invariant.  But since the transformations of the
$\Psi_\lambda$ are unitary, it is easy to form invariant
combinations; one can simply define
\beq
\psi^{B}[A_\bz]
 = \sum_\lambda \bar\Psi_\lambda[B_z]\Psi_\lambda[A_\bz] \ ,
\label{e11}
\eeq
where $B$ is any background field.  The inner product derived from
Chern-Simons theory then gives \cite{EMSS,Ver}
\beq
\langle\psi^{B'}|\psi^{B}\rangle
 = \sum_{\lambda,\lambda'}h^{\lambda\lambda'}
 \bar\Psi_\lambda[B_z]\Psi_{\lambda'}[B'_\bz] \ ,
\label{e12}
\eeq
where $h$ is the diagonal modular invariant.  But \rref{e12} is
simply the partition function for a level $k$ WZW model with
gauge group $G$ coupled to a background gauge field $B$ \cite{Ver}.
It is straightforward to generalize this construction to higher
genus surfaces, and to include vertex operators, which arise
when one incorporates Wilson lines running between the inner
and outer boundaries of the three-manifold \cite{Wita,Kogan,MS}.
A similar approach, introduced in \cite{Witf}, applies
to coset models as well, and it seems likely that amplitudes  for
nondiagonal conformal field theories can be derived from modular
invariant wave functions more complicated than \rref{e11}.

There is more to string theory than conformal field theory, of
course.  It was noted in 1989 that the string theory integration
over moduli space has a natural interpretation in terms of
three-dimensional gravity \cite{CarKog}, and it has recently been
shown that the Liouville term can be derived from the local
Lorentz anomaly in Chern-Simons theory \cite{Cara}.  Notably
missing from the description so far, however, have been the
string theory diffeomorphism ghosts.  The aim of this paper
is to present evidence that these ghosts can be understood
in terms of an overlap integral between initial and final
gravitational wave functions, in close analogy to the conformal
field theory partition function \rref{e12}.

\section{Gravity and Ghosts}

Before proceeding further, we should address a possible
misconception about ghosts and three-dimensional gravity.
The gravitational action is diffeomorphism-invariant, and one
might hope that string theory ghosts would arise from
gauge-fixing the three-dimensional diffeomorphism group.
This is not the case.
In canonical form, the gravitational action is \cite{DJtH}
\beq
S =
  \int dt\int\nolimits_\Sigma
  (\pi^{ab}{\dot g}_{ab} - N^a{\cal H}_a -N{\cal H}) \ ,
\label{e21}
\eeq
where $g_{ab}$ is the spatial metric, $\pi^{ab}$ is its canonical
conjugate, and
\beq
{\cal H} = {1\over \sqrt{g}}(\pi^{ab}\pi_{ab} - \pi^2) -
  \sqrt{g}^{\ \lower1pt\hbox{$\scriptscriptstyle (2)$}}\!R \ ,\qquad
{\cal H}_a = -2\pi_{a\ |b}^{\ b}
\label{e23}
\eeq
are the Hamiltonian and the two-dimensional diffeomorphism
constraints.  It is convenient to choose coordinates in which
$\pi$ depends only on time.  Then just as in string theory,
gauge-fixing the two-dimensional diffeomorphisms generates
a Faddeev-Popov determinant $(\det P^\dagger P)^{1/2}$, where
$P$ is the first-order differential operator that takes vectors
to symmetric traceless tensors.  But
${\cal H}^a = (P^\dagger\pi)^a$, so the path integral over
$N^a$ generates a delta functional $\delta[P^\dagger\pi]$ that
absorbs this determinant.  Similarly, the determinant
coming from gauge-fixing the diffeomorphisms orthogonal to
$\Sigma$ is absorbed by the delta functional generated by
the $N$ integration.  This same phenomenon can be seen
in the Chern-Simons formulation of (2+1)-dimensional
gravity \cite{Witb}, where the path integral gives a ratio of
determinants --- the analytic torsion --- which cancel for
topologies of the form $[0,1]\!\times\!\Sigma$.  If three-dimensional
gravity is to generate the string ghost system, the mechanism
must be more subtle.

Let us now turn to the gravitational analog of the transition
amplitude \rref{e12}.  The action for (2+1)-dimensional gravity
on a manifold $[0,1]\!\times\!\Sigma$ can be viewed as a
Chern-Simons action for the gauge group ISO(2,1) \cite{Witc}.
In the natural (cotangent bundle) polarization, the gravitational
Hilbert space then consists of functions on a set of SO(2,1) holonomies
that characterize the global spacetime geometry \cite{Carc}.  These
states correspond to Heisenberg picture wave functions in ordinary
quantum mechanics.  For our ``thickened string'' picture, however,
we need Schr\"odinger picture wave functions, which contain explicit
information about the  two-dimensional boundaries on which the
string-like excitations live.  The problem is essentially the same
as that of choosing a time-slicing in ordinary gravity, where it
is well known that wave functions determine the time at which
they are defined \cite{Wheeler}.

To understand the choice of time-slicing, it is useful to return to
the canonical metric formulation of (2+1)-dimensional gravity,
where the geometric significance of the phase space variables is
clear.  The phase space of metrics and momenta can be conveniently
divided into two pieces \cite{Moncrief}.  The first consists of
conformal structures on $\Sigma$ and their conjugate momenta,
and has an obvious relevance to string theory.  The second
comprises the conformal factor and its conjugate, the mean
extrinsic curvature $K$ of $\Sigma$.  It is natural to fix a
time-slicing in terms of this latter set of variables; in particular,
Wheeler has argued \cite{Wheela} that York's choice of $K$ as the
time variable \cite{York} is especially attractive from the point
of view of the initial value problem and the variational principle.
Let us make this choice, and examine the Chern-Simons wave
functions for (2+1)-dimensional gravity on a surface of
constant mean extrinsic curvature $K$.

The determination of such wave functions is a difficult problem,
and for most topologies the solution is not known.  If $\Sigma$ has
the topology of a torus, however, the complete answer is given in
reference \cite{Carx}.  Wave functions depend on the mean
curvature $K$ and on two commuting SO(2,1) holonomies
$\exp\{\mu J_2\}$ and $\exp\{\lambda J_2\}$, and are given by
\beq
\psi_{grav}(\mu,\lambda,K)
 = \int_{{\cal F}} {d^2\tau \over(\Imm\tau)^2}
   \left({\mu - \tau\lambda\over\pi (\Imm\tau)^{1/2}K}\right)
   \exp\left\{-{i|\mu - \tau\lambda|^2\over (\Imm\tau) K}\right\}
   \tilde\chi(\tau,\bar\tau) \ ,
\label{e24}
\eeq
where $\tau$ is the modulus of a torus, the integration is over a
fundamental region $\cal F$ for the modular group, and $\tilde\chi$
is any automorphic form of weight $1/2$.  (In contrast to reference
\cite{Carx}, we use the standard notation of string theory,
where $\tau$ denotes the torus modulus rather than the mean extrinsic
curvature.)  The Hamiltonian\footnote{Since gravity is generally
covariant, the ``super-Hamiltonian'' \rref{e23} that generates
translations in coordinate time vanishes identically.  Once a
time-slicing is chosen, however, it is meaningful to discuss a
Hamiltonian that generates motions between slices.} that generates
translations in $K$ is \cite{Carb}
\beq
H = {i\over2}K^{-1}
    \left({\partial\ \over\partial\mu}\mu
    + \lambda{\partial\ \over\partial\lambda}\right) \ ,
\label{e25}
\eeq
and wave functions \rref{e24} are eigenstates of $H$ provided that
$\tilde\chi$ is a Maass form, i.e.,
\beq
 \left( 2i(\Imm\tau){\partial\ \over\partial\tau}-{1\over2} \right)
 \left( 2i(\Imm\tau){\partial\ \over\partial\bar\tau}+{1\over2} \right)
 \tilde\chi(\tau,\bar\tau) = E^2 \tilde\chi(\tau,\bar\tau) \ .
\label{e26}
\eeq
In particular, the ground state is obtained by setting
\beq
\tilde\chi^{(0)}(\tau,\bar\tau,K) = (\Imm\tau)^{1/2}\eta^2(\tau)
\label{e27}
\eeq
in \rref{e24}, where $\eta(\tau)$ is the Dedekind eta function.  A
straightforward calculation shows that the resulting wave function
$\psi_{grav}^{(0)}(\mu,\lambda,K)$ is independent of $K$.

Note that apart from the $\Imm\tau$ factor, the automorphic form
$\tilde\chi^{(0)}$ depends holomorphically on the modulus $\tau$.
This may be a general feature for any genus.  It is possible to
add a gravitational Chern-Simons term to the (2+1)-dimensional
action \rref{e21}, thus introducing dynamical topologically
massive gravitons.  There is then some evidence from perturbation
theory that holomorphicity of the wave functions is a necessary
condition for a positive total Hamiltonian \cite{Kogana}.

The gravitational analog of the inner product \rref{e12} can now be
worked out explicitly.  In particular, for the ground state determined
by \rref{e27}, we find that
\beq
\langle\psi^{(0)}_{grav}|\psi^{(0)}_{grav}\rangle =
\int_{{\cal F}}{d^2\tau\over (\Imm\tau)} |\eta(\tau)|^4 \ .
\label{e28}
\eeq
But this is precisely the diffeomorphism ghost contribution to the
string theory torus partition function.  Combining this result with
\rref{e12} with $B=0$, we find a full transition amplitude of
\beq
\int_{{\cal F}}{d^2\tau\over (\Imm\tau)} |\eta(\tau)|^4
\sum_{\lambda,\lambda'}h^{\lambda\lambda'}
    \bar\Psi_\lambda[0]\Psi_{\lambda'}[0] \ ,
\label{e29}
\eeq
which may be recognized as the full string partition function
for the torus.  Note that the integration over moduli space, with the
correct measure, is forced upon us by the integral over $\lambda$
and $\mu$ in the overlap between the initial and final gravitational
wave functions.  Of course, if the total central charge is nonzero,
an additional Liouville contribution must be included in the string
partition function; but this term also arises in three dimensions, as
a consequence of the local Lorentz anomaly \cite{Cara}.

\section{Conclusion}

We have now seen that the entire one-loop partition function
in string theory can be reconstructed from three-dimensional gravity
and topological field theory.  All of the string theory ingredients
--- the conformal field theory partition function, the vertex
operators, the Liouville action, the diffeomorphism ghosts, and the
integration over moduli space --- have straightforward three-dimensional
interpretations.

Several important issues remain, however.  First, our derivation was
based on a particular choice of time-slicing, which, although natural,
is by no means unique.  Whether the final results depend on this choice
is an open question, related to a fundamental problem in
quantum gravity, that of understanding how general covariance
manifests itself at the quantum level.  Second, our derivation involved
a particular approach to the quantization of (2+1)-dimensional gravity,
based on the Chern-Simons form of the action.  If we had started
instead with metric variables and used a simple prescription for
operator ordering, we would have found wave functions that behaved as
automorphic forms of weight $0$ rather than $1/2$ \cite{Carx,HosNak}.
As yet, we know of no prescription for choosing between these two
approaches to quantum gravity.  Finally, we have not yet succeeded
in generalizing our results to surfaces of genus greater than one, for
which the constant mean curvature slicing becomes quite complicated.

These difficulties are real, and may ultimately defeat the attempt to
describe string theory in terms of three-dimensional topological field
theory.  But in view of the success in reproducing the torus amplitude,
it seems unlikely that the equivalence of the two- and three-dimensional
approaches is merely coincidence.

\vspace{2.5ex}
\begin{flushleft}
\large\bf Acknowledgements
\end{flushleft}

One of us (I.~K.) would like to thank G.~Semenoff and N.~Weiss
for helpful discussions and hospitality at the University of
British Columbia.
S.~C.~was supported in part by the U.S.~Department of Energy
under grant DE-FG03-91ER40674.  I.~K.~was supported in part by
the Natural Sciences and Engineering Research Council of Canada.

\newpage

\newcommand{\NPB}[1]{{\sl Nucl.~Phys.}~{\bf B#1}}
\newcommand{\Ann}[1]{{\sl Ann.~Phys.}~{\bf #1}}
\newcommand{\CMP}[1]{{\sl Commun.~Math.~Phys.}~{\bf #1}}
\newcommand{\PLB}[1]{{\sl Phys.~Lett.}~{\bf B#1}}
\newcommand{\PRL}[1]{{\sl Phys.~Rev.~Lett.}~{\bf #1}}
\newcommand{\PTP}[1]{{\sl Prog.~Theor.~Phys.}~{\bf #1}}
\newcommand{\MPLA}[1]{{\sl Mod.~Phys.~Lett.}~{\bf A#1}}
\newcommand{\IJMP}[1]{{\sl Int.~J.~Mod.~Phys.}~{\bf #1}}
\newcommand{\CQG}[1]{{\sl Class.~Quant.~Grav.}~{\bf #1}}
\newcommand{\PRD}[1]{{\sl Phys.~Rev.}~{\bf D#1}}
\newcommand{\JMP}[1]{{\sl J.~Math.~Phys.}~{\bf #1}}


\begin{thebibliography}{99}

\bibitem{Wita} E.~Witten, \CMP{121} (1989) 351.
\bibitem{DJT} S.~Deser, R.~Jackiw, and S.~Templeton, \PRL{48} (1982)
              975.
\bibitem{Kogan} I.~I.~Kogan, \PLB{231} (1989) 377.
\bibitem{CarKog} S.~Carlip and I.~Kogan, \PRL{64} (1990) 1487;
                 \MPLA{6} (1991) 171.
\bibitem{EMSS} S.~Elitzur, G.~Moore, A.~Schwimmer, and N.~Seiberg,
               \NPB{326} (1989) 108.
\bibitem{MS} G.~Moore and N.~Seiberg, in {\sl Superstrings '89},
             edited by M.~Green et al.~(World Scientific, 1990).
\bibitem{Ver} H.~Verlinde and E.~Verlinde, in {\sl Superstrings '89},
             edited by M.~Green et al.~(World Scientific, 1990).
\bibitem{Witf} E.~Witten, ``On Holomorphic Factorization of WZW and
               Coset Models,'' Institute for Advanced Study preprint
               IASSNS-HEP-91/25 (1991).
\bibitem{Cara} S.~Carlip, \NPB{362} (1991) 111.
\bibitem{DJtH} S.~Deser, R.~Jackiw, and G.~{'t Hooft}, \Ann{152}
               (1984) 220.
\bibitem{Witb} E.~Witten, \NPB{323} (1989) 113.
\bibitem{Witc} E.~Witten, \NPB{311} (1988) 46.
\bibitem{Carc} S.~Carlip, \CQG{8} (1991) 5.
\bibitem{Wheeler} R.~F.~Baierlein, D.~H.~Sharp, and J.~A.~Wheeler,
                  {\sl Phys.~Rev.}~{\bf 126} (1962) 1864;
                  J.~A. Wheeler, in {\sl Relativity, Groups and
                  Topology}, edited by B.~DeWitt and C.~DeWitt
                  (Gordon and Breach, 1964).
\bibitem{Moncrief} V.~Moncrief, \JMP{30} (1988) 2907.
\bibitem{Wheela} J.~A.~Wheeler, \IJMP{A3} (1988) 2207.
\bibitem{York} J.~W.~York, \PRL{28} (1972) 1082.
\bibitem{Carx} S.~Carlip, ``(2+1)-Dimensional Chern-Simons Gravity as a
               Dirac Square Root,'' Davis preprint UCD-91-16 (1991).
\bibitem{Carb} S.~Carlip, \PRD{42} (1990) 2647.
\bibitem{Kogana} I.~I.~Kogan, \PLB{256} (1991) 369.
\bibitem{HosNak} A.~Hosoya and K.~Nakao, \PTP{84} (1990) 739.

\end{thebibliography}
\end{document}